# Control of the structural and magnetic properties of perovskite oxide ultrathin films through the substrate symmetry effect

Jun He,[1,2] Albina Borisevich,[1] Sergei V. Kalinin,[1]
Stephen J. Pennycook,[1,2] and Sokrates T. Pantelides[2,1]

[1]*Oak Ridge National Laboratory, Oak Ridge, Tennessee 37831, USA*
[2]*Department of Physics and Astronomy, Vanderbilt University, Nashville Tennessee 37235, USA*

Perovskite transition-metal oxides are networks of corner-sharing octahedra whose tilts and distortions are known to affect their electronic and magnetic properties. We report calculations on a model interfacial structure to avoid chemical influences and show that the symmetry mismatch imposes an interfacial layer with distortion modes that do not exist in either bulk material, creating new interface properties driven by symmetry alone. Depending on the resistance of the octahedra to deformation, the interface layer can be as small as one unit cell or extend deep into the thin film.

The unique magnetic and electronic properties of perovskite transition metal oxides with $ABO_3$ stoichiometry have attracted broad attention both from fundamental physical perspective and for information storage and logic devices [1,2], with epitaxial films taking center stage as the medium where strain can be explicitly controlled [3-6]. Tailoring the strain can allow stabilization of phases non-existent in the bulk [7-9], induce novel polarization states [10,11], and even enable transition to morphotropic-like states [4,12]. Similarly well recognized are the effects related to the charge at the interfaces arising from chemical or polarization discontinuity [13-15]. The key aspect of all the polarization, chemical, and strain effects is that they act uniformly along the interface, corresponding to zone-center type wavevectors.

However, we note that magnetic, electrical, and structural functionalities of perovskites are also influenced by the deformations and distortions of the octahedral network controlled by zone-boundary phonon modes [16-18]. For magnetic oxides such as ruthenates and manganites, the B-O-B tilt angle between corner-shared octahedra has been shown to affect electron hopping matrix elements and hence the transport properties [19-24]. The deformations of oxygen octahedra driven by collective Jahn-Teller distortions control both metal-insulator transitions and magnetic properties [25]. Recently, first-principles calculations found unexpected ferroelectricity in $PbTiO_3$/$SrTiO_3$ heterostructure, arising from octahedron rotation coupling across the interface [26]. Also, the concept of exchange bias coupling in electrically tunable magnetic devices has been proposed to change the magnetic order laterally at the interfaces between a multiferroic oxide and a ferromagnetic oxide [27]. Thus, effective control of the magnetic and orbital degrees of freedom requires the symmetry control tied to the zone-boundary modes of the material [28].

Calculations on interfaces between dissimilar materials cannot easily unravel the distinct role of symmetry mismatch at the interface from effects arising from the discontinuity in the chemical identity of constituent elements, their valences, and the materials' lattice parameters. On the other hand, Fig. 1 illustrates that symmetry mismatch alone, due to the constraints of the continuity of the octahedral network, imposes a distinct transition layer in which the octahedra undergo additional distortions that are not present in either bulk material. Figure 1a illustrates this point by attempting to connect two materials with different intrinsic octahedral tilts by a single sheet of *undistorted* octahedra with intermediate tilts. Figure 1b shows that the network continuity can be restored by modifications of the interfacial octahedra that can involve any or all of the single-octahedron distortions, such as Jahn-Teller elongation and polar off-centering, as well as additional distortions of the octahedral network, such as new rotational modes. The need for such modifications persists even if one employs a wider transition layer. Clearly, the presence of such symmetry-dictated transition layer has the potential to generate interfacial properties that are distinct from those of the bulk materials.

Another limitation of the accepted approach to the interface calculations lies in employing a superlattice, in which a distinction between the substrate and the thin film cannot be made. In modeling lattice mismatch one can easily impose the planar lattice parameter of the "substrate". Accommodation of the symmetry mismatch in a superlattice, on the other hand, becomes a battle between two thin films. In this Letter, in order to circumvent this difficulty and also to avoid interference from chemical and strain effects, we adopt a model interfacial structure that allows us to unravel the effect of symmetry mismatch at the interface between a substrate and a thin film. We employ an oxide slab of a single material and impose zero out-of-plane tilt in one of the planes. The model implicitly assumes that the substrate has untilted octahedra and is infinitely rigid. In this fashion, we can explore how any particular material accommodates the symmetry mismatch and the concomitant effect on its properties. We envision several scenarios for symmetry relaxation. Depending on the energetics of deforming the individual octahedra, the transition region can be narrow (if the octahedra are easily deformed) or wide (if the octahedral deformation is unfavorable). For the case of a coherent ultrathin film, it is conceivable that the relaxation length could exceed sample thickness. This approach allows us to predict the character and a characteristic length for relaxation of the interface-related octahedral tilt disruptions, and classify different materials based on this behavior.

We performed density-functional theory (DFT) calculations to compare the behavior of two ferromagnetic oxides, $SrRuO_3$ and $La_{0.75}Sr_{0.25}MnO_3$. The results show that the disruption of the tilt system in $SrRuO_3$ is quickly relaxed with full return of the tilts and rotations to the bulk values, while for $La_{0.75}Sr_{0.25}MnO_3$ (in the thickness range used for the calculation) tilts and rotations stabilize at the values substantially different from the bulk. At the same time, local magnetic structure of $La_{0.75}Sr_{0.25}MnO_3$ remains almost unchanged in both strained and unstrained superlattices. For $SrRuO_3$, on the other hand, the local magnetic moment of Ru changes considerably as the changes in octahedral tilts are absorbed. The results suggest that the "rigidity" of the octahedral network is a critical factor in designing oxide thin films and multilayers with desired properties.

$SrRuO_3$ and $La_{0.75}Sr_{0.25}MnO_3$ are chosen as model materials as both are metallic ferromagnetic oxides, often used as bottom electrodes for ferroelectric thin films grown on $SrTiO_3$. These materials have the same tilt system ($a^-a^-c^+$ in Glazer notation) in the bulk [29], albeit very different electronic structure and magnetic properties ($SrRuO_3$ is a narrow-band itinerant ferromagnet with T_c=163 K and La$_{0.75}$Sr$_{0.25}$MnO$_3$ is a double-exchange ferromagnet with T_c=370 K) [30], making them ideal candidates for comparative studies. DFT calculations were performed using a plane-wave basis set and the projector-augmented-wave method as implemented in the Vienna *ab initio* simulation package (VASP) code [31-34]. Calculations for the bulk material (20-atom cells) using a 8x8x8 Monkhorst-Pack k-point mesh and a 500 eV plane-wave cutoff produced structural parameters in good agreement with experimental data. For the 80-atom supercells, a 4x4x1 Monkhorst-Pack k-point mesh and a 400 eV plane-wave cutoff were used. The superlattices were built to contain 16 atomic layers (8 A-O layers and 8 B-O layers) along the z-axis. The orthorhombic structure (space group Pnma) was adopted for SRO and the rhombohedral structure (space group $R\bar{3}c$) was adopted for LSMO. The structures were relaxed using the spin-polarized generalized-gradient approximation for exchange-correlation. The substitution of Sr at La sites in LSMO is simulated using regular arrangements of 3:1 ratio of La to Sr atoms in the appropriate sublattice. To explore the epitaxial strain effects on the tilt and rotation behavior, we have considered the strained superlattice by implicitly constraining the in-plane lattice constant to the bulk lattice parameter of SrTiO$_3$. In this case, SrRuO$_3$ undergoes a compressive strain (~1%) and La$_{0.75}$Sr$_{0.25}$MnO$_3$ undergoes a tensile strain (~2% in a-axis, ~1% in b-axis). We follow the definition of the oxygen octahedron tilting and rotations in reference [24] with respect to different directions. The tilt angle along the z-axis is defined as (180°- Θ)/2, where Θ is the bond angle of B-O-B along z-axis. The rotation angle in x-y plane is defined as (90°-Ω)/2, where Ω is the angle among three oxygen ions between two corner-shared oxygen octahedra in the x-y plane. For unstrained SrRuO$_3$ and La$_{0.75}$Sr$_{0.25}$MnO$_3$, the calculated tilt angles are 10.2° and 7.8°, in good agreement with neutron diffraction measurement (8.6° and 8.3°, respectively) [22,35]. To explore the evolution of octahdral tilts across the interface, the tilt angle is fixed to zero in one plane (virtual interface plane), and the rest of the superlattice is allowed to relax.

We first focus on the tilt angles of $MnO_6$ octahedra along the z-axis in an unstrained $La_{0.75}Sr_{0.25}MnO_3$ superlattice. Figure 2 (a) illustrates the increase of the tilt angle from zero at the virtual interface to the nearly-bulk value at the center of the superlattice. In comparison, the rotation angle becomes larger than the bulk value at the virtual interface plane and decreases gradually towards the bulk value at center of the superlattice. These results indicate that, when the tilts are disabled or suppressed by external perturbations, the out-of-plane rotation becomes the dominant octahedral network distortion in the unstrained $La_{0.75}Sr_{0.25}MnO_3$ superlattice. At the same time, the magnetic moments of Mn ions are found almost unchanged along z-axis from virtual interface plane to the center of the superlattice, consistent with the unchanging geometry of the individual octahedra.

In the case of strained LSMO superlattice [Fig. 2(b)], a similar trend is observed for tilt and rotation angles at the virtual interface plane. However, the tilt and rotation angles now deviate strongly from the strained bulk values. To exclude the finite size effect, we have also performed the calculations on a larger $La_{0.75}Sr_{0.25}MnO_3$ strained superlattice containing 24 atomic layers (12 La(Sr)-O layers and 12 Mn-O layers). In this structure (Fig. 2(c)), we found the tilt angles at the center still deviate from its bulk value (8.0°), reaching saturation at 10.7°. More interestingly, the rotations at the center are effectively suppressed (rotation angles range from 0.3° to 1.8°) by the tensile strain. Similar to the unstrained superlattice, the magnetic moment remains almost constant.

In contrast to LSMO, the local octahedral distortions and magnetic properties of SrRuO$_3$ behave differently when the local symmetry is broken. As shown in Fig. 3(a) and 3(b) for unstrained and strained SrRuO$_3$ superlattices, both tilt and rotation angles reach the bulk values in a distance of three Ru layers. The local magnetic properties in SrRuO$_3$ superlattice are affected by strain effect. Compared in Fig. 3(c) are the local magnetic moments of Ru ions in both unstrained and strained SrRuO$_3$ superlattices in two distortion approaches: one by only setting the tilt angle to zero at the virtual interface plane; another by setting both the tilt and rotation angles to zero at the virtual interface plane. In both approaches, we found a reduction in the magnetic moment across the SrRuO$_3$ superlattice. Especially, when both the tilt and rotation are disabled simultaneously, the magnetism is degraded by ~20% at the virtual interface plane. (Fig. 3(c)) Such degradation shows that the local magnetism of SrRuO$_3$ can be significantly affected in the vicinity of a discontinuity in octahedral tilts.

The possible origins of the difference between the LSMO and SRO behavior include (a) magnetic anisotropy, (b) Jahn-Teller activity, and (c) degree of hybridization between the metal-$d$ and oxygen $p$ orbitals. The orthorhombic SrRuO$_3$ has a magnetic easy axis along the pseudocubic [001] direction while rhombohedral La$_{0.75}$Sr$_{0.25}$MnO$_3$ has an easy axis along the pseudocubic [111] direction [36, 37]. Correspondingly, structural distortion applied in the [001] direction could easily affect SrRuO$_3$, rather than La$_{0.75}$Sr$_{0.25}$MnO$_3$. Secondly, when local symmetry is broken in a strained SrRuO$_3$ superlattice, we observed a decreased Ru-O bond length (about 0.02 Å) along the z-axis in RuO$_6$ octahedra at the virtual interface plane. This change is sufficient to trigger additional Jahn-Teller distortions and increase the magnitude of the t2g splitting, which correspondingly alters the Ru spin configuration [23].

To explore the electronic effects associated with the octahedral tilt control, we studied the electronic structure of both oxides by plotting the projected partial density of states (PDOS) of B-site cation and oxygen in Fig. 4. It is well known that both oxides are half-metallic and the strong hybridization between the B-site cation and oxygen is very important to their magnetic properties. In our study, the calculated spin magnetization is 1.40 μB per Ru in $SrRuO_3$, and 3.37 μB per Mn in $La_{0.75}Sr_{0.25}MnO_3$. As seen from the PDOS, the $La_{0.75}Sr_{0.25}MnO_3$ is purely half-metallic, with only majority spin densities existing at the Fermi level. In contrast, $SrRuO_3$ is a "nearly" half-metallic oxide with both minority spin densities at Fermi level and majority spin densities just below Fermi level. Thus, for $SrRuO_3$, the spin configuration could be easily affected by manipulating the hybridization between Ru d orbitals and oxygen p orbitals through adjusting Ru-O-Ru tiliting angles. However, for the $La_{0.75}Sr_{0.25}MnO_3$, the Mn minority spin density is effectively insulating, as shown in Fig. 4, meaning that it is away from the Fermi level. Thus, the octahedral tilt and rotation distortion in $La_{0.75}Sr_{0.25}MnO_3$ superlattice is insufficient to bring the majority spin band edge closer to Fermi level and the spin magnetization remains unchanged for $La_{0.75}Sr_{0.25}MnO_3$.

Another important aspect of the PDOS of $SrRuO_3$ and $La_{0.75}Sr_{0.25}MnO_3$ is related to the hybridization between B-site cation and oxygen. As seen in Fig. 4, there is strong hybridization in $SrRuO_3$ between Ru d-orbitals and oxygen p orbitals, indicated by the consistent overlap between Ru $t_{2g}$, Ru $e_g$ and O 2p states. As a result, when local symmetry is broken in $SrRuO_3$, these states can "feel" each other and response accordingly. This is the reason why the local magnetism of $SrRuO_3$ is very susceptible to symmetry changes. On the other hand, the hybridization in $La_{0.75}Sr_{0.25}MnO_3$ between the Mn $t_{2g}$, Mn $e_g$ and O 2p states is less pronounced. The Mn $t_{2g}$ and $e_g$ states are essentially completely polarized with a large exchange splitting. Thus, magnetic properties are relatively insensitive to local structural symmetry.

To summarize, we demonstrate the pathway towards control of the electronic and magnetic properties of correlated oxides though the symmetry of the octahedral network of the substrate. The octahedral deformation is an essential element of the interface between materials with dissimilar tilt systems, and, depending on the mode of deformation, can give rise to novel structural, electronic, and magnetic behaviors. The results suggest that "rigidity" of the octahedra should be taken into account when designing thin films and multilayers with desired properties.

Finally, we address the feasibility of experimental implementation of these systems. The static component of interface charge and strain are determined by the lattice parameters and composition mismatch of the material and substrate, and can be manipulated by the field effect in transistor-like structures, chemical adsorption, and remanent polarization. Recently, strain control by piezoelectric substrates has been demonstrated [38]. The direct control of the oxygen tilt network can be achieved through the electric field control of non-equivalent ferroelastic variants (e.g. for [110] oriented rhombohedral perovskites [39]). An even simpler model system is offered by the antiferroelectric materials, where the octahedral tilts and polar order are strongly coupled. We believe that further theoretical understanding of the symmetry-mismatched interfaces between functional oxides will enable control of orbital degrees of freedom for magnetoelectric and other correlated-electron based memory and logic devices [40].

Acknowledgment: This research was sponsored by the Division of Materials Sciences and Engineering, Office of Basic Energy Sciences, U.S. Department of Energy, by DOE grant

**Figures**

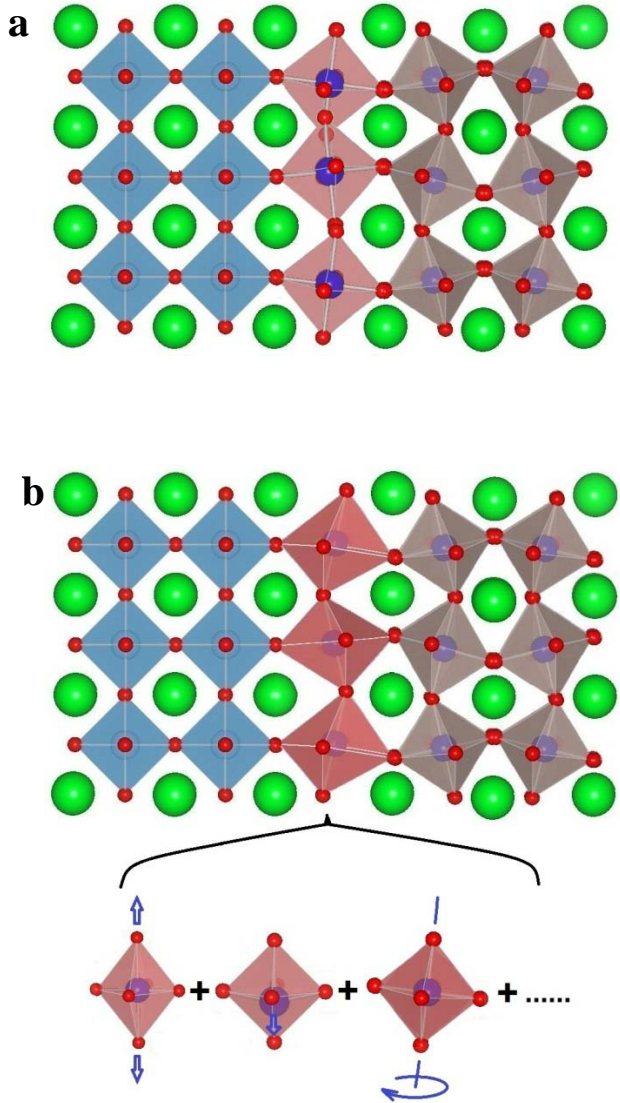

FIG. 1. (Color online) Schematics of the oxygen octahedral deformations at the interface between two perovskite oxides. (a) When slabs of untilted (blue octahedra) and tilted (beige octahedra) perovskites are "connected" via an interfacial layer (pink) of undistorted octahedra of intermediate tilt, the continuity of the corner-sharing network is disrupted (note disconnected vertices in the interface plane). (b) The continuity can be restored by distorting the interfacial octahedra. Shown under the bracket are the possible components of the resulting distortion, such as Jahn-Teller elongation, polar off-centering, and additional rotational modes. Oxygen atoms are in red, A-site atoms are in green, B-site atoms are in grey and blue.

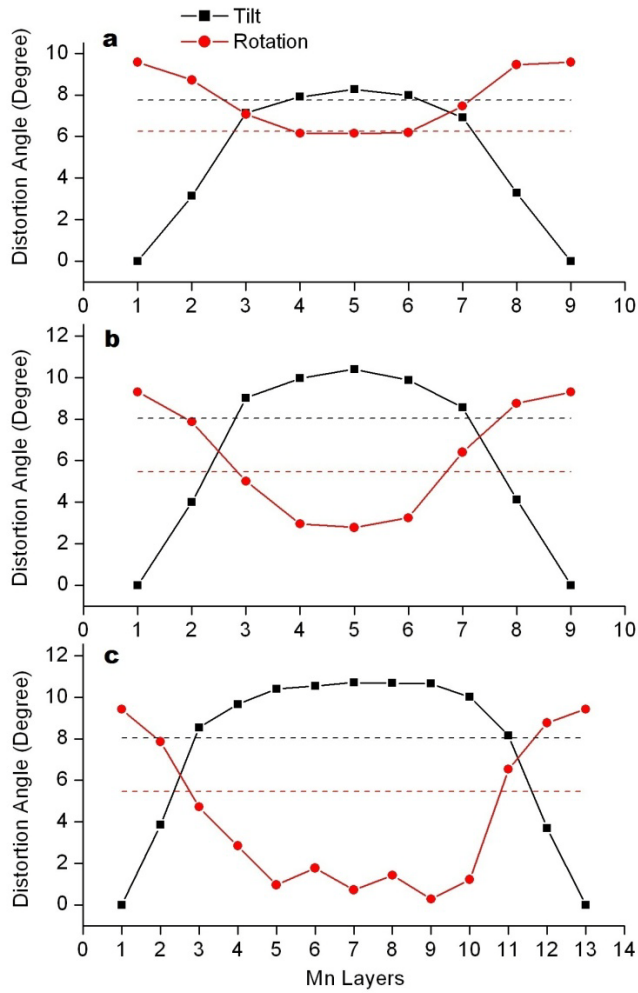

**Fig. 2** Octahedral tilt and rotation angles for $La_{0.75}Sr_{0.25}MnO_3$ as a function of Mn layers along z-axis. The tilt angle is set to zero at the boundary. The dashed lines indicate the values of corresponding bulk tilt angles (black) and rotation angles (red). (a) Unstrained in-plane lattice (b) Strained in-plane lattice. (c) Strained in-plane lattice for a larger supercell.

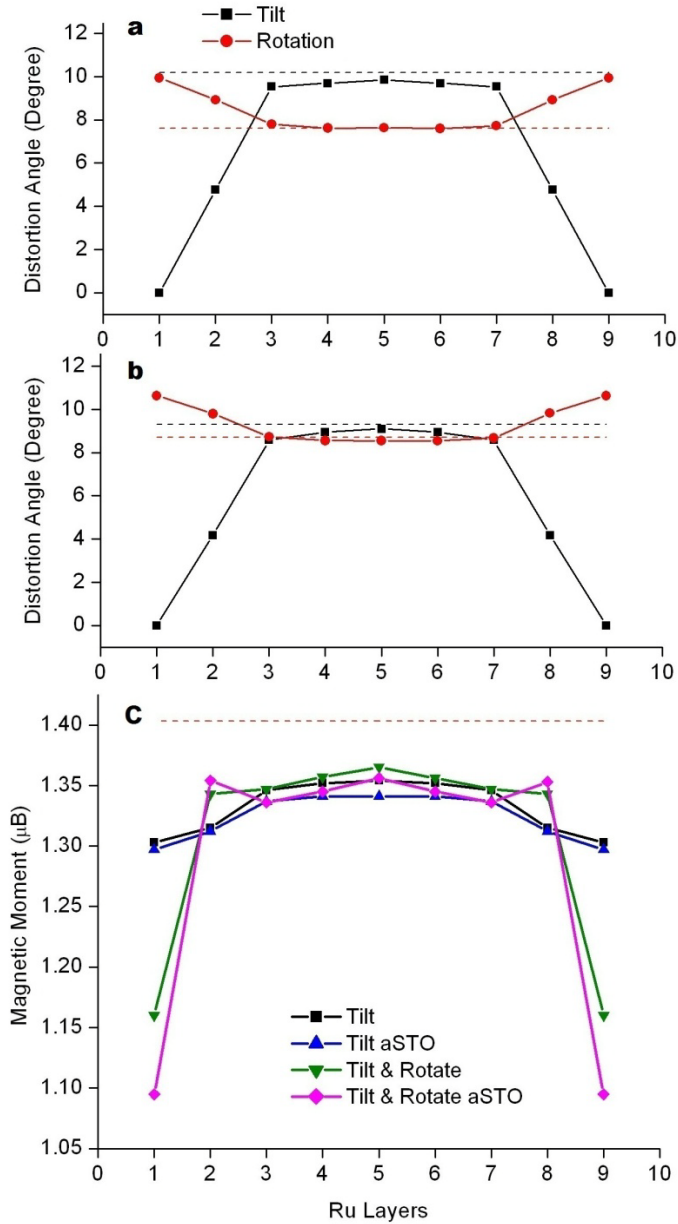

**Fig. 3.** (color online) Octahedral tilt and rotation angles as a function of Ru layers along the z-axis and magnetic moment profile of Ru atoms as a function of Ru layers along the z-axis in the SrRuO$_3$ superlattice. The tilt angle is set to zero at the boundary. The dashed lines indicate the values of corresponding bulk tilt angles (black) and rotation angles (red). (a) Unstrained in-plane lattice (b) Strained in-plane lattice. The lattice constants are fixed to the SrTiO$_3$

substrate and introduce a compressive strain of about 1% in both a-axis and b-axis. (c) The tilt angle is set to zero at the boundary for unstrained (black square) and strained (blue triangle) SrRuO$_3$. Both tilt and rotation angles are set to zero at the boundary for unstrained (green triangle) and strained (purple diamond) SrRuO$_3$. The dashed lines indicate the values of corresponding bulk magnetic moment of Ru in unstrained SrRuO$_3$.

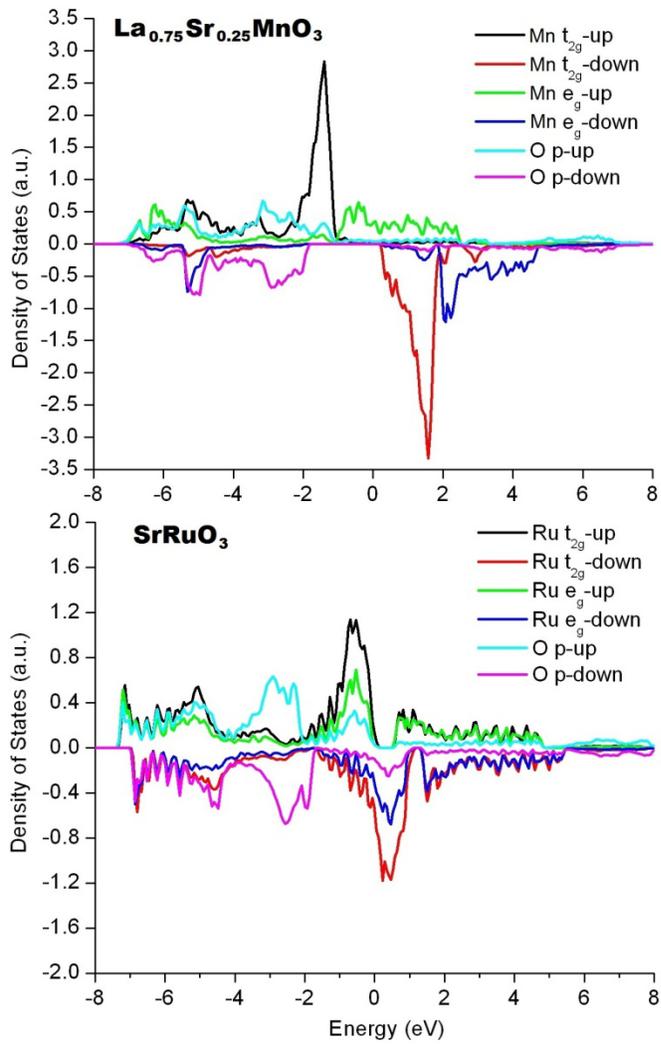

**Fig. 4.** Partial density of states of Mn and O ions in unstrained bulk $La_{0.75}Sr_{0.25}MnO_3$ (a); and Ru and O ions in unstrained $SrRuO_3$ (b). The majority spin is plotted as up spin and the minority spin is plotted as down spin.